\documentclass[aps,prb,superscriptaddress,showpacs,floatfix,twocolumn,amsmath]{
revtex4-1}
\usepackage{graphicx}
\usepackage{bm,amsmath,amssymb,mathrsfs,dcolumn}
\usepackage{subfigure}
\usepackage{units}
\usepackage{hyperref}
\usepackage{mhchem}
\usepackage{mathtools}
\usepackage{wrapfig}
\usepackage{multirow}
\usepackage[usenames]{color}
\usepackage[export]{adjustbox}
\hypersetup{pdfborder=0 0 0,colorlinks=true,citecolor=blue,linkcolor=blue}
\newcommand{\angstrom}{\mbox{\normalfont\AA}}

\begin{document}

\title{The Finite 
Temperature Structure of the MAPbI$_3$ 
Perovskite: Comparing Density Functional Approximations and Force Fields to Experiment}
\date{\today}

\author{Jonathan Lahnsteiner}
\email{jonathan.lahnsteiner@univie.ac.at}
\affiliation{Faculty of Physics, University of 
Vienna, Computational Materials Physics, Sensengasse 8/12, 1090 Vienna, 
Austria}
\author{Georg Kresse}
\affiliation{Faculty of Physics, University of 
Vienna, Computational Materials Physics, Sensengasse 8/12, 1090 Vienna, 
Austria}
\author{Jurn Heinen}
\affiliation{Van 't Hoff Institute for Molecular Sciences, University of 
Amsterdam, Science Park 904, 1098XH, Amsterdam, The Netherlands}
\author{Menno Bokdam}
\email{menno.bokdam@univie.ac.at}
\affiliation{Faculty of Physics, University of 
Vienna, Computational Materials Physics, Sensengasse 8/12, 1090 Vienna, 
Austria}

\begin{abstract}
Determining the finite temperature structure of the hybrid perovskite 
MAPbI$_3$ is a challenge for both experimental and 
theoretical methods. A very 
powerful computational method that can resolve the atomic structure is molecular dynamics (MD). 
The resulting structure depends on the density functional approximation (DFA) in the case of \textit{ab initio} MD and the force field in classical MD. We compare the structure between 250~K and 400~K obtained with different DFAs and force fields in one consistent manner. The symmetry of the PbI$_3$ framework is analyzed as well as the relative ordering of the neighboring organic molecules inside the framework. The distribution function of the molecules is used to map out an effective energy surface for the rotation of a single molecule. This surface is accurately modeled by a pair of cubic harmonics. Available experimental data in literature are discussed and compared to the structure obtained with the different methods. The spread in these data is still too large to uniquely determine the method that 'best' 
describes the perovskite, however promising candidates and outliers have been identified.

\end{abstract}

\maketitle

\section{Introduction}
The stark rise in photovoltaic efficiency of hybrid perovskites and its 
potential use in very thin, flexible and cheap solar 
cells has spiked interest in the condensed matter community. The most successful perovskite until now is the 
Methylammonium Lead-Iodide (MAPbI$_3$). The interesting dynamics of the molecules in this ionic crystal at finite temperatures and their hypothesized influence on the electronic properties\cite{Frost:nanol14,Neukirch:nanol16} warrant thorough investigation of its atomic structure. This 
perovskite system has already been studied with experimental techniques such as X-ray diffraction (XRD)\cite{Stoumpos:ic13,Baikie:jmca:13,Weller:cc105,Whitfield:sr16,Franz:crat2016}, dielectric spectroscopy\cite{Onoda-Yamamuro:jpcs:92,Govinda:jpcl17}, infra-red (IR) spectroscopy\cite{Bakulin:jpcl15}, nuclear magnetic resonance (NMR) spectroscopy\cite{Wasylishen:ssc85,Franssen:jpcl2017} or quasi-elastic-neutron scattering (QENS)\cite{Onoda-Yamamuro:jpcs90,Chen:pccp15}.
From the theoretical side, extensive density functional theory (DFT) 
studies have been performed. With \textit{ab initio} 
molecular dynamics (MD) simulations, insight into the finite temperature structure has been obtained\cite{Carignano:jpcc15,Mosconi:com2015,Lahnsteiner:prb16,Druzbicki:jpcl16,Carignano:jpcc2017}.
Because of the necessary dimensions of the supercell to avoid spurious 
interactions and the limitations in computational resources,
classical force fields\cite{Mattoni:jpcc15,Mattoni:jp2017,Handley:pccp2017} have been designed and parametrized 
relying on DFT calculations. 
Furthermore, the complexity of the material has inspired the construction of 
model Hamiltonians\cite{Frost:aplm14,Leguy:natc15,Simenas:jpcl17,Tan:acsel2017}, describing the interactions of Methylammonium (MA) dipoles 
on a grid, in order to determine the essential underlying 
physics.

Out of all these 
studies, a general picture arises. The MAPbI$_3$ perovskite adopts a 
stable tetragonal unit cell at room temperature (RT), while 
allowing for considerable movement of the Pb-I framework as well as the MA 
molecules. The accuracy of the different experimental methods are, however, 
limited, allowing for different interpretations at the level of the 
PbI octahedra and the MA molecules orientations. Firstly, many experiments propose rotations of 
MA molecules at finite temperature 
\cite{Onoda-Yamamuro:jpcs:92,Bakulin:jpcl15,Leguy:natc15,Whitfield:sr16,Govinda:jpcl17}, but the measured
speed of this process differs as well as the related energy barriers. At 
$300$~K, the QENS measured rotational energy 
barrier for the molecular 
carbon nitrogen axes (CN-axes) is $13.5$~meV and has a reorientation 
time of $13.8$~ps in the tetragonal MAPbI$_{3}$ structure\cite{Leguy:natc15}.
From dielectric permittivity measurements Onoda-Yamamuro 
\textit{et.al.}~\cite{Onoda-Yamamuro:jpcs:92}
obtained $101$~meV for the rotational barrier of the CN-axes.
In a second work by Onoda-Yamamuro~\textit{et.al.} using QENS, a 
rotational barrier of $27$~meV
and a reorientation time of $1.0$~ps was found \cite{Onoda-Yamamuro:jpcs90}.
Wasylishen~\textit{et.al.} used $^{2}H$ and $^{14}N$ NMR spectroscopy to 
determine 
the values of the rotational barrier of the CN-axes as 
$52$~meV\cite{Wasylishen:ssc85}.
The experiment was carried out at $330$~K for the tetragonal MAPbI$_{3}$ 
phase and $0.463$~ps was obtained as the
reorientation time. IR spectroscopy measurements 
were 
done by Bakulin~\textit{et.al.} at 300~K and they obtained 
a reorientation time of
$3$~ps\cite{Bakulin:jpcl15}.
Chen~\textit{et.al.} used QENS analysis on cubic MAPbI$_{3}$ at $350$~K and obtained $70$~meV for the 
rotational barrier and $2.7$~ps for the reorientation time\cite{Chen:pccp15}. These experimental data are summarized in Table~\ref{T1:exp_data}.
Secondly, there are several experimental studies that obtain the molecular 
orientations. Measurements of the second harmonic generation signal at different delay time by Govinda~\textit{et.al.}, rule out the possibility of a transient ferroelectric state of the molecules under photoexcitation\cite{Govinda:jpcl17}. Whitfield~\textit{et.al.} proposes, 
based on synchrotron X-ray powder diffraction and time of 
flight neutron diffraction analysis, that the MA molecules in the 
cubic phase are dynamically disordered but prefer to be aligned along 
the principal cubic axes \cite{Whitfield:sr16}.
In the work by Weller~\textit{et.al.} powder neutron diffraction analysis 
is used to determine that the MA molecules in the 
cubic \ce{MAPbI_{3}} phase also align
the MA molecules along the cubic axes $\langle100\rangle$ \cite{Weller:cc105}.
On the other hand, the work of Franz~\textit{et.al.}, where X-ray powder diffraction, neutron 
powder diffraction and synchrotron X-ray diffraction was used, reports that the molecules are oriented along the $\langle221\rangle$ 
directions of the cubic lattice\cite{Franz:crat2016}.

\begin{table}
 \caption{Experimental measurements of the molecular reorientation time 
($\tau$) and rotational barrier as obtained by various 
experimental techniques.}
 \label{T1:exp_data}
   \begin{ruledtabular}
   \begin{tabular}{c c c c c}
    T [K] & Phase & time [ps] & barrier [meV] & method \\
    $300$K  &  tetragonal  &  13.8   &  13.5         & QENS \cite{Leguy:natc15} \\
      -     &  tetragonal  &   -     & 101           & Permitivity \cite{Onoda-Yamamuro:jpcs:92} \\
    $330$K  &  tetragonal  &  0.463  &  52           & NMR \cite{Wasylishen:ssc85} \\
    $300$K  &  tetragonal  &   3     &   -           & IR \cite{Bakulin:jpcl15} \\
    $350$K  &  cubic       &  2.72   &  94           & QENS \cite{Chen:pccp15} \\
    $300$K  &  tetragonal  &  5.37   &  78           & Permitivity \cite{Poglitsch:jcp87} \\
    $300$K  &  tetragonal  &  1.0    &  27           & IR \cite{Onoda-Yamamuro:jpcs90} \\
    $289$K  &  cubic       &   -     &  108          & NMR \cite{Xu::JPS1991} \\
   \end{tabular}
  \end{ruledtabular}
\end{table}

In the present work the atomic structure of the MAPbI$_{3}$ perovskite is studied in a wide temperature range around the tetragonal-cubic phase transition by two computational methods: \textit{ab initio} molecular dynamics and classical molecular dynamics. Multiple different density functional approximations (DFAs) and force fields are analyzed in one consistent manner.
Note that the choice of the DFA is a very critical one. In Ref.~\onlinecite{Bokdam:prl17} we have 
reported on this issue by assessing the accuracy of different density 
functionals compared to accurate many body perturbation theory 
calculations. In short, the SCAN density functional was 
shown to outperform most standard and van der Waals corrected density 
functionals. Here, we report on our attempts to find the 'best' 
method by comparing calculated physical parameters to the experimentally 
obtained equivalents. Out of the molecular dynamics data, we determine the structure of the PbI framework by mapping the orientation of the PbI$_6$ octahedra onto distribution functions. Hereafter, a similar procedure is applied to the MA molecules. By a Boltzmann inversion of the MA distribution function, we calculate 
an effective energy surface for molecular rotation within the Pb-I perovskite cage. The corrugation energy for the molecules is then defined as the difference between the highest and lowest energy on this surface. Note that the experimental rotational barrier will be lower than this corrugation energy, because multiple reorientation pathways with a lower barrier exist on the surface. We show that this surface can be accurately described by a pair of cubical harmonics. By averaging over the autocorrelation functions of each individual molecule we obtain the reorientation times of the MA molecules. Besides the local orientation of the molecules, also the relative orientation with respect to the nearest neighbors is analyzed.  We will show that the 
current level of accuracy and consistency in and between experiments is 
insufficient to assign a single 'best' density functional or force field for finite 
temperature structure calculations of the hybrid perovskites.

The paper is organized as follows: in Section~\ref{sec:compm} the applied computational methodology is explained. The structural analysis methods are presented in Section~\ref{sec:ana}. The results of the calculations are presented in Section~\ref{sec:res}. These results are discussed and compared to experiment in Section~\ref{sec:dis}. In the last Section ~\ref{sec:conc}, we summarize the outcomes in conclusions.


\section{Computational methods}
\label{sec:compm}

\subsection{DFT}\label{dft1}
All the DFT calculations in this paper were carried out using
VASP\cite{Kresse:prb93,Kresse:cms96,Kresse:prb96} (Vienna ab initio simulation package).
The electronic minimization was computed within the projector augmented
wave (PAW)\cite{Kresse:prb99} formalism. The Brillouin zone was sampled 
using 4 k-points which are set according 
to the cubic symmetry of the $2\times2\times2$ supercell and include the $\Gamma$, R, M and Z points.
For the broadening of the one electron levels we use Gaussian smearing with $\sigma=0.05$~eV. 
To treat the electron-electron exchange-correlation energy we use several 
different density functional approximations. The following DFAs were 
selected to represent a wide range of commonly used functionals: 
the local density approximation (LDA)\cite{Kohn:pr65}, Perdew-Burke-Ernzerhof 
(PBE)\cite{Perdew:prl96}, 
optPBE (vdW DFT functional based on PBE)\cite{Klimes:prb11}, PBE with Grimme D3 vdW correction (PBED3)\cite{Grimme:jcp10},
PBE revised for solids (PBEsol)\cite{Perdew:prl08}, Tkatchenko-Scheffler with vdW correction (TS)\cite{Tkatchenko:prl09}
and, a new meta-GGA, strongly constrained appropriately normed 
(SCAN)\cite{Sun:prl15}. The energy cutoff for the size of the plane wave basis 
set was set to $250$~eV for
all functionals except of SCAN, for which 300~eV was used. SCAN uses 
the kinetic energy gradient of the wave function and therefore more 
accurate orbitals are required. These relatively small plane-wave 
basis sets are possible, because only the outer shell electrons are 
treated explicitly for every atom. For Pb the ($6$s$^{2}6$p$^{2}$), for I 
($5$s$^{2}5$p$^{5}$), C($2$s$^{2}2$p$^{2}$), N($2$s$^{2}2$p$^{3}$) and H 
($1$s$^{2}$) orbitals are included in the valence shell of the pseudo-potential.
The one electron orbitals are determined and converged until the stopping 
criterion ($10^{-4}$~eV) for the self consistent electronic cycle is reached. The effect of spin-orbit coupling on the structure is negligible\cite{Perez:jpcc17} and therefore was not included.

\subsection{ab initio Molecular Dynamics}
After the electronic ground state of a system has been obtained, an \textit{ab initio} 
MD trajectory can be calculated by computing the forces acting 
on the ions using the Hellmann-Feynman theorem. The Born-Oppenheimer surface is explored by propagating the ions forward 
in time with a small time step. The temperature was maintained during the MD simulations
by a Langevin thermostat \cite{Allen:book91,Bussi:pre2007}. The time step was set to $10$~fs for all 
MD calculations except when the SCAN functional was used; here a time step of 
$8$~fs seconds was used.
The smaller time step for the SCAN functional was used to get higher precision orbitals within
an acceptable number of electronic minimization steps.
These relatively large time steps are possible 
because the internal degrees of freedom of the MA molecules were constrained.
The rotation of the NH$_3$ relative to the 
CH$_3$ group over the C$-$N bond was not constrained, thereby allowing the $C_{3}$
rotation of the molecule.
To constrain the remaining internal degrees of freedoms, bond lengths and 
certain bond angles, the Shake algorithm \cite{Bucko:jcp2005,Baker:jcp1999}
was applied. This algorithm creates virtual forces for the desired degrees of 
freedom to maintain their values during the simulations. 
Additionally, increased hydrogen masses of $4$~a.u where used. The hereby reduced motion in the system makes the Shake algorithm converge faster. The increased hydrogen mass lowers the 
reorientation time by $10$\%. This was checked by 
comparison of two MD simulations with the same starting structure. Keep in mind that the choice of the starting 
structure potentially has a more sizable influence on the
reorientation times than the heavier hydrogen masses\cite{Lahnsteiner:prb16}.  
As a further validity check, we compared a constrained MD simulation (Shake algorithm 
applied) with an unconstrained MD, the 
hydrogen masses were fixed to the same value but the time step was decreased 
to $2$~fs. This test made clear that the internal 
molecular degrees of freedom do not influence the here presented results at all. 
The total simulation times are tabulated for each functional and temperature in Table~\ref{T2:MDtime}. To improve the sampling efficiency, a parallel tempering scheme\cite{Franz:crat2016} was used additionally for optPBE, PBED3, TS and SCAN. For each DFA, four MD trajectories run in parallel with their thermostats set to 250, 300, 350 and 400~K. During the run, the temperatures can switch between the trajectories depending on the energy differences, see Ref.~\onlinecite{Bokdam:prl17} for more details. Parallel tempering was used for the van der Waals corrected functionals, because the van der Waals forces 
hinder or slow down the dynamics. The dynamics using the other DFAs is so 
fast that it is not necessary to use the parallel tempering approach. The 
obtained probability distributions are the same for fixed temperature MDs  
(provided sufficiently long trajectories are used) as when the 
parallel tempering scheme is used. This was checked explicitly for the PBEsol 
and PBE functionals. The disadvantage of parallel tempering is that the self correlation times ($\tau$) for the MA molecules cannot be extracted from the autocorrelation function. Therefore $>$200~ps long NVT trajectories were calculated additionally.

\begin{table}
 \caption{Length (in ps) of the continues \textit{ab initio} MD trajectories and between brackets the parallel tempering times, for pseudo-cubic $2\times2\times2$ MAPbI$_3$ supercells. The additional parallel tempering calculations are used for the calculation of the distribution functions.}
 \label{T2:MDtime}
 \begin{ruledtabular}
 \begin{tabular}{c c c c c }
    DFA:    & 250~K & 300~K & 350~K & 400~K \\
    LDA     & 600  & 401  & 400  & 400 \\
    PBE     & 200  & 200  & 200  & 200 \\
    PBEsol  & 200  & 200  & 200  & 200 \\
    optPBE  & 375 (450)  & 200 (450)  & 200 (450)  & 200 (450) \\
    PBED3   & 200 (200)  & 296 (200)  & 600 (200)  & 617 (200) \\
    TS      & 500 (750)  & 200 (750)  & 200 (750)  & 200 (750) \\
    SCAN    & 230 (264)  & 220 (264)  & 204 (264)  & 207 (264) 
 \end{tabular}
 \end{ruledtabular}
\end{table}

\subsection{Structural Models Considered}

The central objects of study in this work are $2\times2\times2$ supercells of MAPbI$_3$ with 
pseudo-cubic symmetry. The imposed lattice constants are adapted from 
single crystal XRD analysis\cite{Stoumpos:ic13}. The lattice 
constants are set to $a=b=6.312$~\angstrom{} and $c=6.316$, therefore our supercells have a small tetragonal distortion of $0.06$\%. The volume of a unit cell is $\text{V}=251~\text{\AA}^{3}$. According to Wahl et al., the properties of perovskite systems are very sensitive to the unit cell volume\cite{Wahl:prb2008}. Our initial calculations showed a sensitive dependence of the molecular orientation on the unit cell volume. Therefore, additional MD simulations of cubic unit cells with a smaller volume of $\text{V}=245$~$\text{\AA}^{3}$ and a lattice constant of $a=6.257$~\angstrom{} were calculated for comparison. This is the equilibrium volume for tetragonal MAPbI$_3$ using the SCAN functional at room temperature\cite{Bokdam:prl17}.

The considered MAPbI$_{3}$ unit cell has 
$3\cdot12=36$ atomic degrees of freedom. After applying the constraints
to the MA molecules, only $19$ degrees of freedom remain. One standardized frozen MA geometry is used for all MD calculations even though different DFAs are used. This approximation was checked by calculating the optimal bond lengths and dipole moments of the MA molecule in vacuum for different functionals. As can been seen in Table~\ref{molecules}, the differences in bond lengths between the various functionals are minor. The intrinsic dipole moment is between 2.1 and 2.2~D for all functionals, except for the LDA which contracts the C-N bond length too much and thereby reduces the dipole to 2.0~D. The PBEsol optimized molecular geometry is used for all simulations in this work. To obtain a starting configuration the molecules are placed in the PbI$_{3}$ 
framework with random starting orientations. Hereafter the MD trajectory is calculated and at every time step the geometry is analyzed.

\begin{table}
 \caption{Molecular bond lengths (in $\text{\AA}$) and dipole moments $\mu$ 
(in Debye) for the relaxed ground state structure of the Methylammonium 
molecule in vacuum as obtained for different DFAs.}
\begin{ruledtabular}
 \begin{tabular}{c c c c c }
  DFA: & CN & \ce{CH} & \ce{NH} & $\mu$\\

  LDA    & 1.487 & 1.070 & 1.063 & 2.00 \\
  
  optPBE & 1.510 & 1.096 & 1.062 & 2.12 \\
  
  PBE    & 1.511 & 1.097 & 1.058 & 2.17 \\
  
  PBED3  & 1.523 & 1.098 & 1.059 & 2.13 \\
  
  PBEsol     & 1.500 & 1.100 & 1.061 & 2.17 \\
  
  SCAN   & 1.508 & 1.093 & 1.058 & 2.17 \\

  TS     & 1.511 & 1.097 & 1.058 & 2.17
  \end{tabular}
 \end{ruledtabular}
  \label{molecules}
\end{table}

\subsection{Classical Molecular Dynamics}

Classical molecular dynamics calculations were performed with RASPA 2.0  
Molecular Software Package for Adsorption and Diffusion in (Flexible) 
Nanoporous Materials\cite{Dubbeldam:ms2016}. The forces were calculated according to two different 
force fields parametrized based on DFT data by Mattoni~\textit{et.al.}~\cite{Mattoni:jpcc15}
and another one by Handley \textit{et.al.}~\cite{Handley:pccp2017}. The force fields were implemented in the RASPA code. The simulations were done in the NVT ensemble.
To maintain the temperature a Nose-Hoover thermostat was used. The time step during the MD was set to $0.5$~fs and every $5^{th}$ step a sample was taken which is equivalent to
taking a sample every $2.5$~fs. Before starting sampling, an equilibration period 
of $5$~ps was calculated. In total $2,000,000$ steps were computed without counting the equilibration 
steps. This corresponds to a total sample number of $40,000$ steps and a total 
simulation time (without equilibration time) of $1,000$~ps. For the van 
der Waals and the Coulomb interactions the cutoff radii  were set to 12~\AA{} which is 
slightly smaller than half of the box size. The electrostatic interactions are calculated with 
the Ewald summation method and the precision is set to $10^{-9}$. 
The lattice vectors are the same as for the \textit{ab initio} molecular 
dynamics, but larger $4\times4\times4$ super-cells were used and the constraints on the molecules were omitted.



\section{Structure Analysis method}
\label{sec:ana}

\begin{figure}
 \centering
   \includegraphics[width=0.8\linewidth]{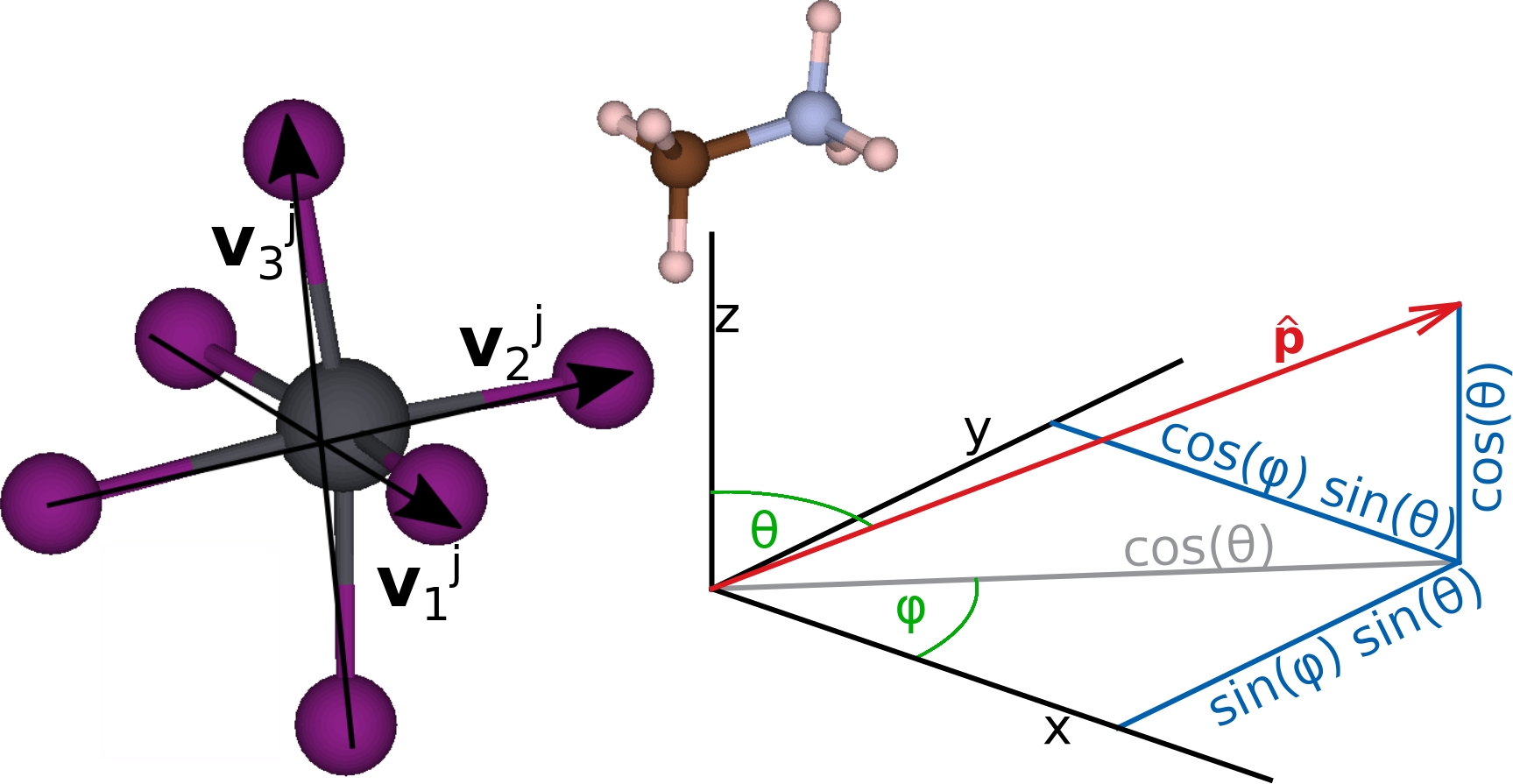}
 \caption{The $j^{th}$ lead atom with surrounding iodine atoms. $\hat{\mathbf{v}}_{\nu}^{j}$ denotes
          the iodine-iodine vectors defined in positive Cartesian direction $\nu$. The PbI framework creates a cubical cavity in which the MA molecule reorients. This movement is described by the unit vector $\mathbf{\hat{p}}$ parallel to the C-N axis and is expressed in spherical coordinates ($\theta,\phi$). }
 \label{frame_polar_sketch1}
\end{figure}

\subsection{Effective potential energy surface by Boltzmann inversion}
The spherical angular distribution function $\rho$($\theta_{i}$,$\varphi_{i}$) of the vector $\mathbf{\hat{p}}$ 
describing the \ce{C-N} axes of the MA molecule is extracted from the finite temperature MD trajectories.
As illustrated in Figure~\ref{frame_polar_sketch1}, $\theta$ denotes the angle of $\mathbf{\hat{p}}$ with the $\mathbf{\hat{z}}$ direction of the coordinate 
frame, and $\varphi$ is the angle in the $\mathbf{\hat{x}\hat{y}}$-plane,
chosen to be zero when aligned with the $\mathbf{\hat{x}}$-axes. After assigning the molecules
for all time frames to the directions ($\theta_{i}$,$\varphi_{i}$), the cubic 
symmetry of the lattice is invoked to down-fold these spherical distributions 
onto a single octant. The same procedure was used in 
Ref.~\onlinecite{Lahnsteiner:prb16}, where it was shown that the dynamical correlation between the neighboring molecules is low. Here, we approximate the spherical angular distribution function $\rho$($\theta_{i}$,$\varphi_{i}$)
to be independent of the other molecules as well as of the cage. Therefore we can average over all molecules, $\rho(\theta,\varphi)=\frac{1}{N}\sum_i\rho(\theta_{i},\varphi_{i})$. The probability distribution $\rho(\theta,\varphi)$ for a particular molecular orientation can be written as,
\begin{equation}
 \rho(\theta,\varphi) = \frac{e^{-\beta E( \theta , \varphi ) } }{ 
\int_{0}^{\pi}\int_{0}^{2\pi} e^{-\beta E(\theta,\varphi) }d\varphi 
d\theta}=\frac{e^{-\beta E( \theta , \varphi ) }}{\mathcal{Z}},
 \label{bi1}
\end{equation}
with $\beta=\frac{1}{k_{B}T}$, $k_{B}$ the 
Boltzmann constant, $T$ the temperature and ${\mathcal{Z}}$ the partition 
function in the independent molecule picture.  The potential energy surface of the molecules is calculated from 
the down folded polar distributions by a Boltzmann inversion. The effective energy surface $E(\theta{},\varphi{})$ denotes the energy of a molecule being oriented along a certain direction. Taking the natural logarithm of Eq. (\ref{bi1}) and 
solving for $E( \theta , \varphi )$ we obtain
\begin{equation}
 E( \theta , \varphi ) = -k_{B}T\left( \ln ( \rho( \theta , \varphi )) - \ln( \mathcal{Z}) \right ).
 \label{bi2}
\end{equation}
This inversion technique determines the energy surface up to an 
unknown additive constant. This suffices, since we are only interested in 
energy differences. In this work, the energy surfaces are defined such that the minimum value of the 
energy surface is set to zero as shown in Figure \ref{e_surf_orig_350K_fit}.

\subsection{Modelling the potential energy surface}
To compare the energy surfaces obtained with the 
different methods, a small set of cubic harmonics is fitted to 
the energy surfaces. Since pseudo-cubic cells are used, cubic 
harmonics posses the correct symmetry as long as there is no sizable symmetry breaking.
The applied cubic harmonic functions can be written as
\begin{equation}
 f_{1}( \alpha_{x} , \alpha_{y} , \alpha_{z} ) = \alpha_{x}^{2}\alpha_{y}^{2} + 
                  \alpha_{y}^{2}\alpha_{z}^{2} + \alpha_{x}^{2}\alpha_{z}^{2},
 \label{ch1}
\end{equation}
and 
\begin{equation}
 f_{2}( \alpha_{x} , \alpha_{y} , \alpha_{z} ) = \alpha_{x}^{2}\alpha_{y}^{2}\alpha_{z}^{2}.
 \label{ch2}
\end{equation}
Here the $\alpha_{i}$ are the directional cosines of the \ce{C-N} 
bond vectors ($\mathbf{\hat{p}}$). The directional cosines 
$\alpha_{i}$ are calculated by 
\begin{equation}
\nonumber
 \alpha_{x} = \mathbf{\hat{x}}\cdot\mathbf{\hat{p}}, \ \ \alpha_{y} = \mathbf{\hat{y}}\cdot\mathbf{\hat{p}},\ \ \alpha_{z} = \mathbf{\hat{z}}\cdot\mathbf{\hat{p}}.
\end{equation}
The cubic lattice and the PbI$_3$ cage impose symmetry on the polar distributions and energy surfaces of the molecules. The essential symmetry can be captured by a superposition of these cubic harmonics
\begin{equation}
 \begin{split}
     E_{fit}(\mathbf{\hat{p}}) =   C_{1} \left [ \alpha_{x}^{2}\alpha_{y}^{2} + 
                  \alpha_{y}^{2}\alpha_{z}^{2} + \alpha_{x}^{2}\alpha_{z}^{2} \right ] 
                  + C_{2} \alpha_{x}^{2}\alpha_{y}^{2}\alpha_{z}^{2}+C_{3}.
    \label{fit_func}
 \end{split} 
\end{equation}
The potential energy surface imposed by different methods can hereby be reduced to three
parameters $C_{1}, C_{2}$ and $C_{3}$.  The optimal parameters are calculated by least-square fittings of Eq.~(\ref{fit_func}) to the energy surfaces obtained from MD simulations. This step was done with the DGLS routine, part of the Lapack library\cite{LAPACK:1999}. In this way the noise, which remains even after the long MD trajectories, is averaged out.

The symmetry of the cubic harmonics terms of Eq.~(\ref{fit_func}) and the high symmetry orientations of the molecule are graphically represented on the top of Figure~\ref{e_surf_orig_350K_fit}. Depending on the ratio between $C_1$ and $C_2$, the maximal corrugation energy ($\Delta{}E$) and the minimal corrugation energy ($\Delta{}E_{\rm min}$) can be calculated. In this model, the height of the potential barrier along the path from the $\mathbf{\hat{x}}$ orientation over the face-diagonal $\frac{1}{\sqrt{2}}(\mathbf{\hat{x}}+\mathbf{\hat{z}})$ to the $\mathbf{\hat{z}}$ orientation is
\begin{equation}
\Delta{}E_{x\rightarrow{}xz}=\frac{C_1}{4}.
\label{barxxz}
\end{equation}
The barrier from the $\mathbf{\hat{x}}$ orientation over the room-diagonal $\frac{1}{\sqrt{3}}(\mathbf{\hat{x}}+\mathbf{\hat{y}}+\mathbf{\hat{z}})$ to the $\mathbf{\hat{z}}$ orientation is
\begin{equation}
 \Delta{}E_{x\rightarrow{}xyz}=\frac{C_1}{3}+\frac{C_2}{27}.
 \label{barxxyz}
\end{equation}
The barrier from the face-diagonal orientation over the room-diagonal to another face-diagonal is
\begin{equation}
 \Delta{}E_{xz\rightarrow{}xyz}=\frac{C_1}{12}+\frac{C_2}{27}.
 \label{barxzxyz}
\end{equation}
Lastly, when $C_1$ and $C_2$ have opposite sign (and $C_2$ is sufficiently large) a saddle point $\mathbf{\hat{s}}(C_1,C_2)$ appears on the surface. The barrier from the face-diagonal orientation over $\mathbf{\hat{s}}$ to another face-diagonal ($\Delta{}E_{xz\rightarrow{}s}$) was determined numerically.

\subsection{Reorientation times of the molecules}
For the calculation of the reorientation times ($\tau$) of the MA molecules, the autocorrelation
function averaged over all molecules is computed,
\begin{equation}
 r( t - t_{0} ) = \frac{1}{N}\sum_{i=1}^{N} 
\mathbf{\hat{p}}_{i}(t_{0})\cdot\mathbf{\hat{p}}_{i}(t).
 \label{auto1}
\end{equation}
Here $N$ denotes the number of molecules in the cell and 
$\mathbf{\hat{p}}_{i}(t_{0})$,
$\mathbf{\hat{p}}_{i}(t)$ are the \ce{C-N} unit vectors at time $t_0$ and $t\ge{}t_{0}$, respectively. 
The autocorrelation function is obtained by computing a time average of 
Eq.~(\ref{auto1}) over $N_{T}$ different starting times,
\begin{equation}
 \left < r( t ) \right > = \frac{1}{N_{T}}\sum_{j=0}^{N_{T}-1}r( t - \Delta t 
\cdot j ),
 \label{auto2}
\end{equation}
with $\Delta t = \frac{T}{2(N_{t}-1)}$ for $N_{T}>1$. In this work $\Delta{}t$ was set to $100\times$ the MD time step, resulting in one pico-second. Hereafter, the reorientation
times of the molecules are extracted by fitting $ f( t ) = e^{-\tau{}t}$ to the autocorrelation function. Since the obtained functions $ \left < r( t ) \right > $ are not simple exponentials, the fit was made in two steps. First, a fit was made including the entire data range, from this we determined the half-life period $\tau$ followed by a second fit in the range $[0,\tau]$. This procedure was iterated until convergence was reached (usually after two iterations). 

\subsection{Molecular ordering pattern}
A statistical measure for the relative orientation of the molecules is obtained by calculating 
the dot product between its neighboring molecules,
\begin{equation}
d(t) = \frac{1}{N}\sum_{i=1}^N\sum_{j\in 
n.n.} \hat{\mathbf{p}}_{i}(t)\cdot\hat{\mathbf{p}}_{j}(t).
 \label{rel_o}
\end{equation}
This quantity is calculated for every time step $(t)$ of the MD trajectory and
binned into a histogram, thereby forming the probability distribution for 
the relative molecular orientations. This distribution is computed separately for the first, second and third nearest neighbor molecules $j$.

\section{Results}
\label{sec:res}


\subsection{Structure of the PbI$_3$ framework}\label{frame_sec}

In the MAPbI$_3$ perovskite every lead atom is bonded to six nearest neighbor 
iodine atoms forming an octahedron with lead in the center. As illustrated in 
Figure~\ref{frame_polar_sketch1}, four iodines 
lie in the $xy$ plane and two along the $z$ axes. The Pb-I bond is a mixture of a covalent and 
an ionic bond. Considering the $\text{MA}^{+}$ molecule as one undividable element in the $ABX_3$ perovskite structure, the Pb-I bond constitutes the largest contribution to the cohesive energy 
of the crystal. The $\text{MA}^{+}$ molecules are only weakly bonded to the framework and 
are electronically decoupled from it. A crucial step for the theoretical description of a material is 
the choice of an applicable approximation, in this case a density functional or a classical force field.
To asses the different methods we have extracted the \ce{I-I} bond vectors ($\mathbf{v}^{j}_{\nu}$) from the MD trajectories. The \ce{I-I} bonds are chosen such that in the center of the bonds a lead atom is located, see Fig.~\ref{frame_polar_sketch1}.
The vectors $\mathbf{v}^{j}_{\nu}$ are rescaled to unity as denoted 
by $\hat{\mathbf{v}}_{\nu}^{j}$. Next they are transformed to spherical
coordinates $(\theta^{j}_{\nu},\phi^{j}_{\nu})$ and assigned to a two-dimensional probability distribution. This representation describes the rotation of an octahedron as a whole. Another possibility is to define
lead-iodine bond vectors and transform them to spherical coordinates. This representation would also be capable to capture the internal deformations of an octahedron, but has not been pursued in the present work.

\begin{figure}[b]
 \centering
 \includegraphics[width=1\linewidth]{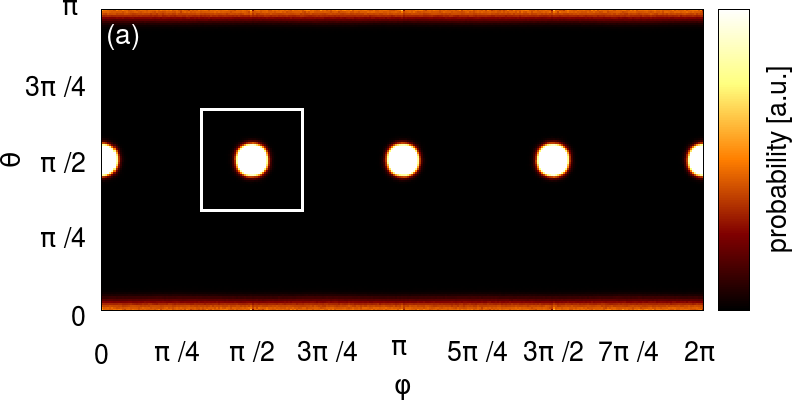}
 \includegraphics[width=0.85\linewidth]{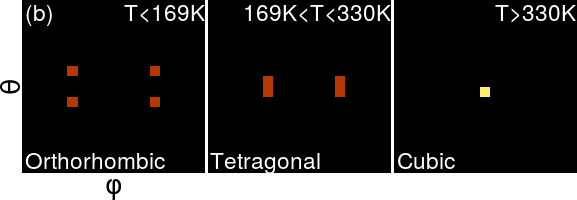}
 \caption{ 
 Probability distribution for angles 
$(\theta,\phi)$ corresponding to the \ce{I-I} bond vectors extracted from (a) an MD trajectory calculated with the Mattoni force field at 350~K and (b) typically optimized orthorhombic, tetragonal and cubic structures from Ref.~\onlinecite{Lahnsteiner:prb16}. (In (b) only the square area around the $\hat{\mathbf{y}}$ direction is shown.)}
 \label{frame_polar_sketch}
\end{figure}

As an example of the probability distribution of the I-I vectors, the distribution obtained from 
a classical MD trajectory calculated with the Mattoni force field\cite{Mattoni:jpcc15} is shown in 
Figure~\ref{frame_polar_sketch}(a). It shows that the I-I vectors point along the Cartesian $x,y,z$-directions forming a cubic framework. The projection 
of spherical coordinates to a rectangular plot smears out the intensity of the 
'north and south poles' ($\theta \pm \pi$) over a large area. Computing the 
integral around the $x,y$ and $z$ directions shows that all are equally 
likely. Because almost perfectly cubic supercells (``pseudo-cubic'' from here on) are used, the 
probability distribution does not differ between the three different orthogonal \ce{I-I} vectors associated with an octahedron.
We will therefore zoom-in (throughout this work) on a single square around the $y$-direction (see Fig.~\ref{frame_polar_sketch}(a)) and show the $\{\nicefrac{\pi}{3}\le\phi\le\nicefrac{2\pi}{3},\nicefrac{\pi}{3}
\le\theta\le\nicefrac{2\pi}{3}\}$ window of the plot. This zoom-in functions as a descriptor for the PbI framework are illustrated in Fig.~\ref{frame_polar_sketch}(b). Here we have analyzed typical structures obtained by DFT PBEsol optimization of experimental orthorhombic, tetragonal and cubic unit cells\cite{Lahnsteiner:prb16}. The XRD experiments indicate that the iodine-iodine vectors form a characteristic tilting pattern in the $xy$-plane depending on the temperature\cite{Whitfield:sr16,Weller:cc105}. Qualitatively, this tilting pattern is the same along the $x$ and $y$ directions. When the temperature is increased, a transition from four spots in the orthorhombic phase, to two in the tetragonal phase and finally to one spot in the cubic phase is observed. In the orthorhombic structure neighboring octahedra are clockwise/counter-clockwise rotated in the $xy$-plane and tilted to the left/right in the $z$-direction. This results in four spots. The tilting in the $z$-direction is removed (or reduced) in the tetragonal phase and two spots remain. In the cubic structure the rotations are also removed and an, on average, simple cubic framework remains as indicated by just one spot. 
We will use these three characteristic distributions as a guideline to determine which structure is dominating in the MD simulations. But we need to be aware that such an assignment remains very cursory for the following reasons:
(i) all simulations were performed for cubic or close to cubic cells and the tetragonal distortion was absent.
(ii) This implies that tetragonally ordered structures can temporarily develop with the $x$, $y$ or $z$ axis standing out. Such patterns might prevail over a substantial simulation time, but can dynamically change in the course of the simulations from say $x$ to $y$ order. (iii) The exact nature of the tetragonal phase is still unknown. For instance, it could be either related to a static tetragonal order, or what we believe is more likely, it could be a highly dynamical phase that fluctuates between orthorhombic local minima.\cite{Bokdam:prl17} (iv) Related to these issues, our simulation do
not allow us to define a proper order parameter for the tetragonal phase, for instance, because the $c/a$ ratio was kept fixed.

\subsection{Temperature dependence of the PbI$_3$ framework}\label{frame_sec_temperature}

\begin{figure*}[!t]
 \centering
 \includegraphics[width=1\linewidth]{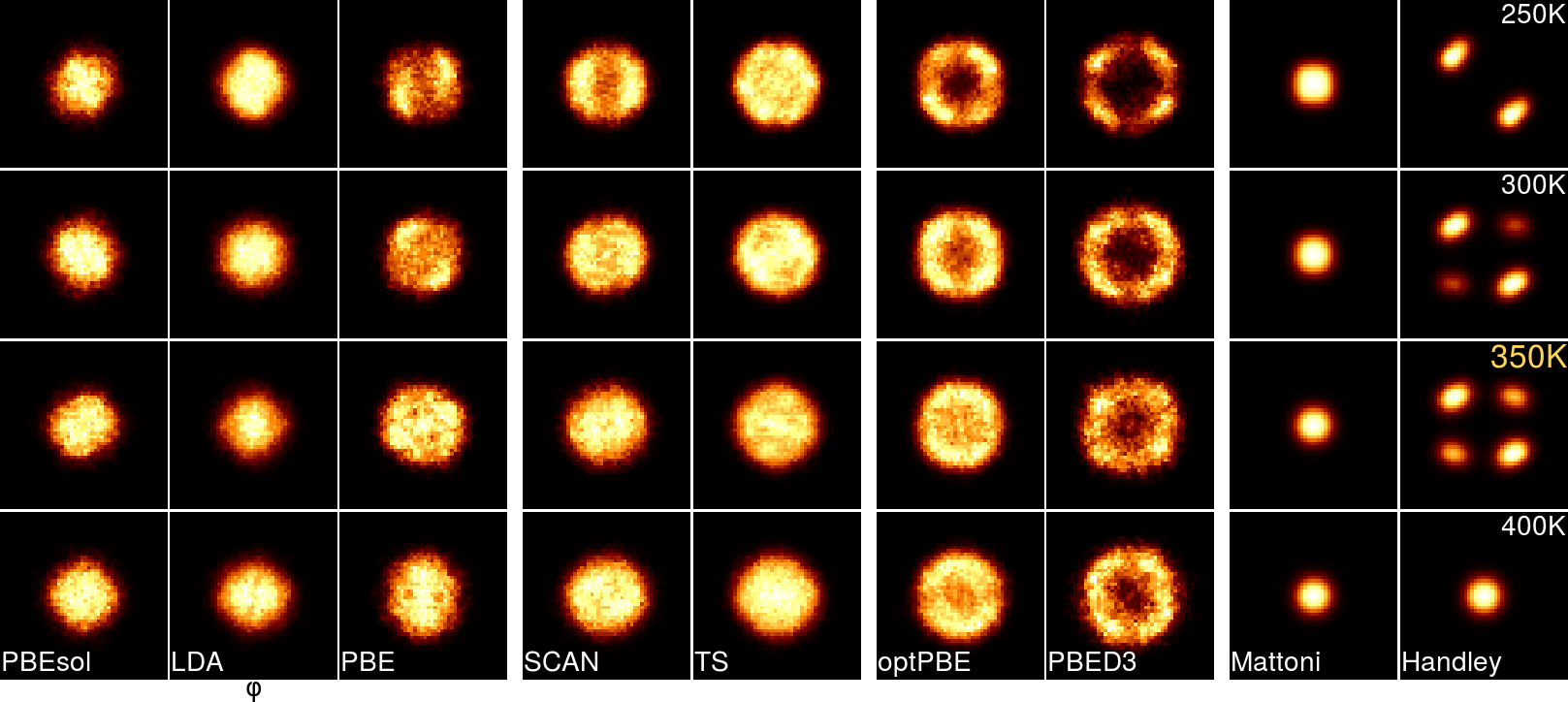}
 \caption{Probability distribution of angles $(\theta,\phi)$ of the \ce{I-I} bond vectors for the different methods in the temperature range from $250$~K to $400$~K. The center of each plot denotes the $\hat{\mathbf{y}}$ direction $(\pi/2,\pi/2)$. The lower left and the upper right corner of each square are the $(\pi/3,\pi/3)$ and $(3\pi/2,3\pi/2)$ directions, respectively. The brighter the color the more frequent the orientation occurs.}
 \label{frame_polars_T}
\end{figure*}

Figure \ref{frame_polars_T} shows the probability distributions of the I-I vectors for the 
different computational methods at temperatures between 250~K and $400$~K. We first focus on the results obtained at 350~K as shown in the third row. The different methods display distinct patterns. Since the tetragonal to cubic transition temperature is $\sim330$~K, the simlations should show signs of the cubic phase and the distribtion function must be symmetric  around the cubic axes. The XC-functionals PBEsol, LDA, PBE, SCAN, TS and the Mattoni force field show a clear maximum along the cubic axes. The radius of the circles differ, indicating differences in the stiffness of the PbI framework. The smaller the radius, the stiffer the framework. On average, 
the \ce{I-I} bonds  appear to be perfectly aligned with the Cartesian axes. PBE-D3 and to some extent optPBE show a slightly different pattern, indicating a framework consisting of alternating rotated and tilted PbI octahedra. The distribution of the Handley force field shows four distinct sharp spots. This indicates a very rigid PbI framework with tilted clockwise/counter clockwise orientations of the octahedra in the $xy$-plane.

To study to what extent these methods can describe the cubic to tetragonal phase transition, the distributions at 250, 300 and 400~K have also been calculated. The columns in Fig.~\ref{frame_polars_T} show the temperature dependence of the orientational probability distributions of the I-I bonds for every method. As explained before,
we expect some change but due to the deficiencies enumerated above the assignment needs to be rather cursory.
The methods are arranged in four groups according to the stiffness of the framework and the shape of the distributions. 
If the method is able to describe the cubic to tetragonal transition, we expect to see a reduction of symmetry of the distribution ($1\rightarrow2$~spots) at some temperature 
between 400 to 250~K. 
The first group consists of PBEsol, LDA and PBE. This group shows deformed circularly shaped distributions around the principal axes and a stiffer framework compared to the remaining XC-functionals. The second group containing SCAN and TS shows softer framework bonds and circular distributions for temperatures above $250$~K. The third group consists of optPBE and PBE-D3, they tend to avoid the directions along the principal axes. This behavior is more pronounced with decreasing temperature, especially for PBE-D3. The force fields make up the fourth group and show the stiffest description of the framework. Mattoni shows circular spots for all temperatures, whereas the Handley force field shows a ``transition'' not observed with any of the DFAs.
Comparing Fig. \ref{frame_polars_T} with Fig. \ref{frame_polar_sketch}(b) one can interpret the SCAN and PBE distributions at 250~K and the optPBE at 300~K as tetragonal like patterns. An orthorhombic like 
pattern can be seen in the four spots of the PBE-D3 and optPBE distributions at $250$~K.


\subsection{Potential energy surface of the molecules}
\begin{figure*}[!t]
 \includegraphics[width=1\linewidth,center]{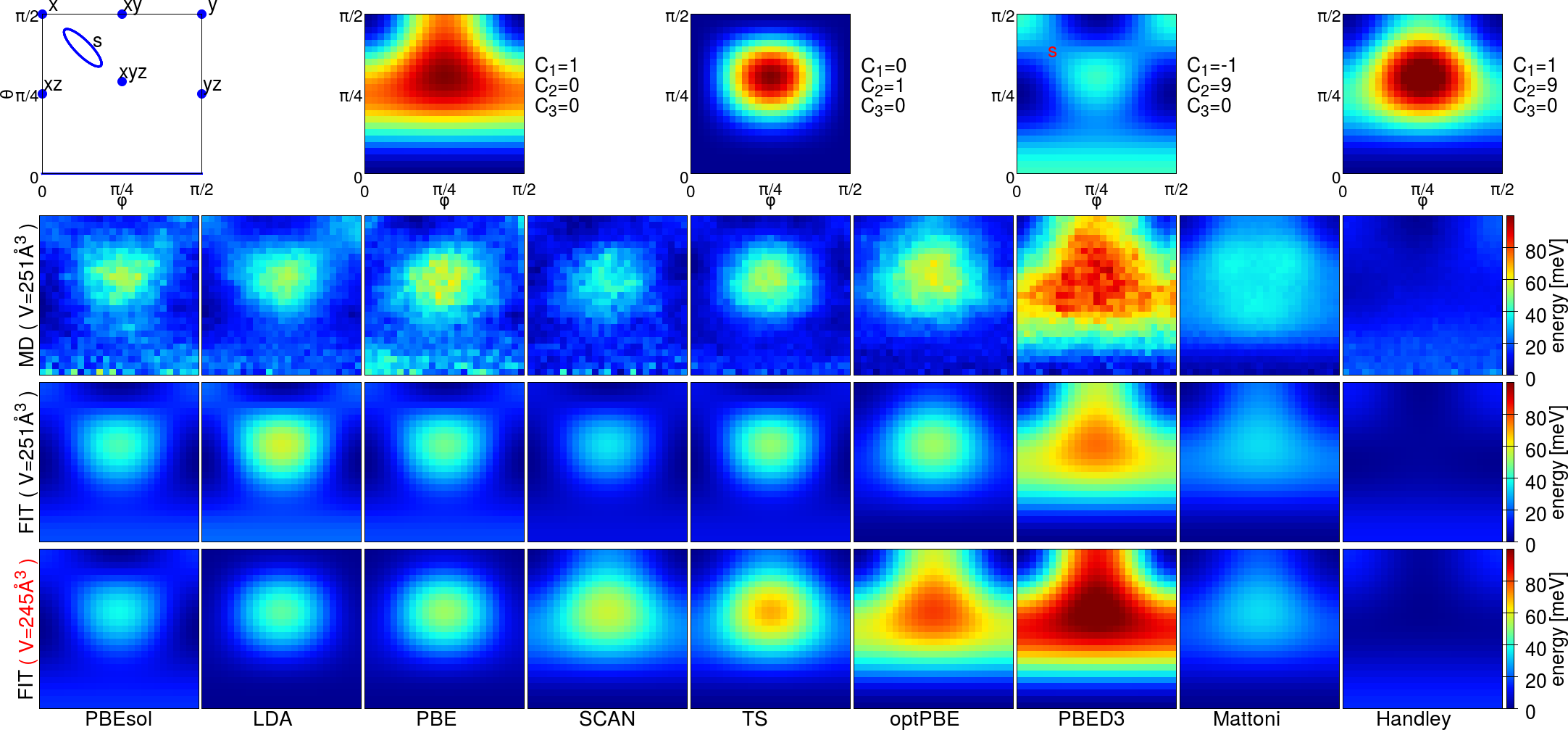}
 \caption{Effective potential energy surfaces of the $\text{MA}^{+}$ molecules for different DFAs and force fields at 350~K. \textbf{$1^{\text{st}}$ row}: (From left 
to right) The high symmetry orientations, cubic harmonic functions $f_{1}$, $f_{2}$ and two linear combinations of them.
 \textbf{$2^{\text{nd}}$ row:} Energy surfaces at $V=$~$\text{251}$~$\text{\AA}^{3}$ obtained by Boltzmann inversion of distribution function $\rho(\theta,\varphi)$.
 \textbf{$3^{\text{rd}}$ row:} Fitted energy surfaces at $V=$~$\text{251}$~$\text{\AA}^{3}$.
\textbf{$4^{\text{th}}$ row:} Fitted energy surfaces at $V=$~$\text{245}$~$\text{\AA}^{3}$.}
 \label{e_surf_orig_350K_fit}
\end{figure*}

After having characterized the structural degrees of freedom of the PbI framework, we turn our attention to the $\text{MA}^{+}$ molecules. The orientation of all molecules are extracted from the MD trajectories, and the potential energy surface as a function of the molecular orientation is calculated by a Boltzmann inversion of the distribution function. The raw potential energy surfaces are shown in Figure~\ref{e_surf_orig_350K_fit} ($2^{\text{nd}}$~row). Cubic harmonic functions (Eq. \ref{fit_func}) are fitted to them and the resulting model potential energy surfaces are shown in the $3^{\text{rd}}$ row. The model consists of two cubic harmonic terms ($C_{1}$ and $C_{2}$), with characteristic symmetries as illustrated in the legend of Fig.~\ref{e_surf_orig_350K_fit} (top) and an offset ($C_{3}$). Depending on the sign, the $C_{1}$ term suppresses or enhances the face diagonal ($xz$,$xy$,$yz$) and room diagonal ($xyz$) orientations of the molecule in the cubic PbI cage. The $C_{2}$ term only affects the room diagonal ($xyz$) orientations. A linear combination of the two accurately describes the original MD surfaces as is verified by calculating the Pearson-correlation coefficient between the two, see Tables~\ref{corr_coeffs1}~\&{}~\ref{corr_coeffs2} of the Appendix. As rule of thumb, correlation values above $0.9$ indicate that the simulated data is well 
described. All obtained fitting parameters for the different methods and temperatures are documented in Tables~\ref{Tfit1} and ~\ref{Tfit2} of the Appendix.

We focus now on the model results obtained at 350~K as shown in Figure \ref{e_surf_orig_350K_fit}. At this temperature the system is cubic and the molecules can preform complete reorientations since the minimal corrugation energy ($\Delta{}E_{\rm min}$) on the potential energy surface is of the order of $\sim\nicefrac{1}{2}k_{b}T$. Depending on the symmetry of the surface, $\Delta{}E_{\rm min}$ can be any of the four barriers expressed by Eqs.~(\ref{barxxz}-\ref{barxzxyz}) and $\Delta{}E_{xz\rightarrow{}s}$. The darker blue areas in the energy surface plots indicate the more likely molecular orientations. At this temperature all orientations occur during the MD, but some occur more often than others. The PBEsol functional tends to orient the molecules along the face diagonals $xy$, $xz$ and $yz$ as does the LDA, PBE, SCAN and the TS functional. This symmetry of the potential energy surface is captured by the sign of $C_1$. Since $C_1<0$, the barrier between an axes orientation and the face diagonal ($\Delta{}E_{x\rightarrow{}xz}$) is negative, as can be seen in Eq.~(\ref{barxxz}). In the second category with $C_1>0$, to which PBE-D3, optPBE as well as the Mattoni force field belong, the MA molecules are primarily directed along the principal axes $x$,$y$ and $z$. The Handley force field is an outlier, it allows all MA orientations except for those along
the principal axes and even allows the MA molecules to be oriented along the room diagonals $xyz$.

Table~\ref{timesbar} shows the maximal ($\Delta {E}$) and minimal ($\Delta{}E_{\rm min}$) corrugation energies for the different XC-potentials and temperatures. $\Delta {\rm E}$ is defined as the difference between the extrema of the fitted energy surfaces. $\Delta {\rm E}$ is slightly decreasing with temperature, making it easier for the molecules to undergo reorientations. At the same time, the framework becomes more flexible at higher temperatures. At $350$~K and $400$~K the corrugation energies for the different DFAs differ only slightly. At 350~K all, with the exception of two, methods show $\Delta{}E_{\rm min}\sim13$~meV and $\Delta{}E\sim43$~meV. The PBE-D3 functional constrains the molecules more and the Handley force field shows a significantly flatter surface. The temperature has no effect on the dominant symmetry of the energy surface, the preferred orientations remain the same.

The volume of the supercell has a large effect on the potential energy surface. Supercells with only a $2.4\%$ smaller volume (245~$\text{\AA}^{3}$) can already have different preferred orientations for the molecules, as shown in the last row of Fig.~\ref{e_surf_orig_350K_fit}. By contracting the cell, LDA and PBE show a slight preference for the molecules to be oriented along the principal axes. The SCAN and the TS functional are now aligning the molecules along the principal axes. The other methods do not switch the preferred orientations for the molecules. The potential energy surface for the SCAN ($V=$~245~$\text{\AA}^{3}$, $\nicefrac{c}{a}=1$) calculation, is very similar to the tetragonal ($V=$~245~$\text{\AA}^{3}$, $\nicefrac{c}{a}=1.01$) calculation presented in Ref.~\onlinecite{Bokdam:prl17}. The effect of the tetragonal distortion is that a higher fraction of the molecules align along the $x$, $y$ axes compared to the $z$ axes\cite{Bokdam:prl17}.


\begin{table*}
\caption{Reorientation times $\tau$ (in ps) and the minimal ($\Delta {\rm E}_{\rm min}$) and maximal ($\Delta {\rm E}$) corrugation energies (in meV) of the effective potential energy surfaces of the MA molecules in MAPbI$_3$.}
\label{timesbar}
\begin{ruledtabular}
\begin{tabular}{ c ccc ccc ccc ccc }
T&\multicolumn{3}{c }{250~K}&\multicolumn{3}{c }{300~K}&\multicolumn{3}{c }{350~K}&\multicolumn{3}{c }{400~K}\\
         & $\tau$ & $\Delta {E}_{min}$ & $\Delta {E}$&$\tau$&$\Delta {E}_{min}$ & $\Delta {E}$&$\tau$&$\Delta {E}_{min}$& $\Delta {E}$ & $\tau$&$\Delta {E}_{min}$ & $\Delta {E}$ \\
PBEsol    &   18     & 14 &     50     &     8.1     & 15 &     50     &     7.2     & 14 &     43     &     3.8     & 14 &     42    \\ 
LDA     &     12       & 17 &     48       &     5.5     & 17 &     48     &     4.1     & 13 &     43     &     2.4     & 14 &     43    \\ 
PBE     &     19     & 12 &     55     &     7.3     & 15 &     54     &     7.2     & 14 &     47     &     4.2     & 12 &     45    \\ 
SCAN     &     9        & 10 &     48     &     5.2     & 10 &     41     &     3.5     & 8  &     35     &     2.2     & 9  &     34    \\ 
TS     &     99     & 11 &     61     &     14     & 10 &     59     &     14     & 9  &     50     &     5.1     & 10 &     47    \\ 
optPBE     &     52       & 20 &     67     &     31     & 20 &     62     &     10     & 14 &     51     &     6.0     & 13 &     44    \\ 
PBED3     &     **     & 56 &     78     &     47     & 48 &     74     &     19     & 55 &     76     &     9.6     & 49 &     75    \\ 
Mattoni     &     1.9     & 17 &     38     &     1.3     & 19 &     33     &     1.0     & 23 &     34     &     0.7     & 25 &     34    \\ 
Handley     &     1.6     & 19 &     19     &     1.4     & 16 &     16     &     0.9     & 12 &     12     &     0.5     & 14 &     14    \\ 
\end{tabular}
\end{ruledtabular}
\end{table*}

\begin{figure}[!b]
 \centering
 \includegraphics[width=8.6cm]{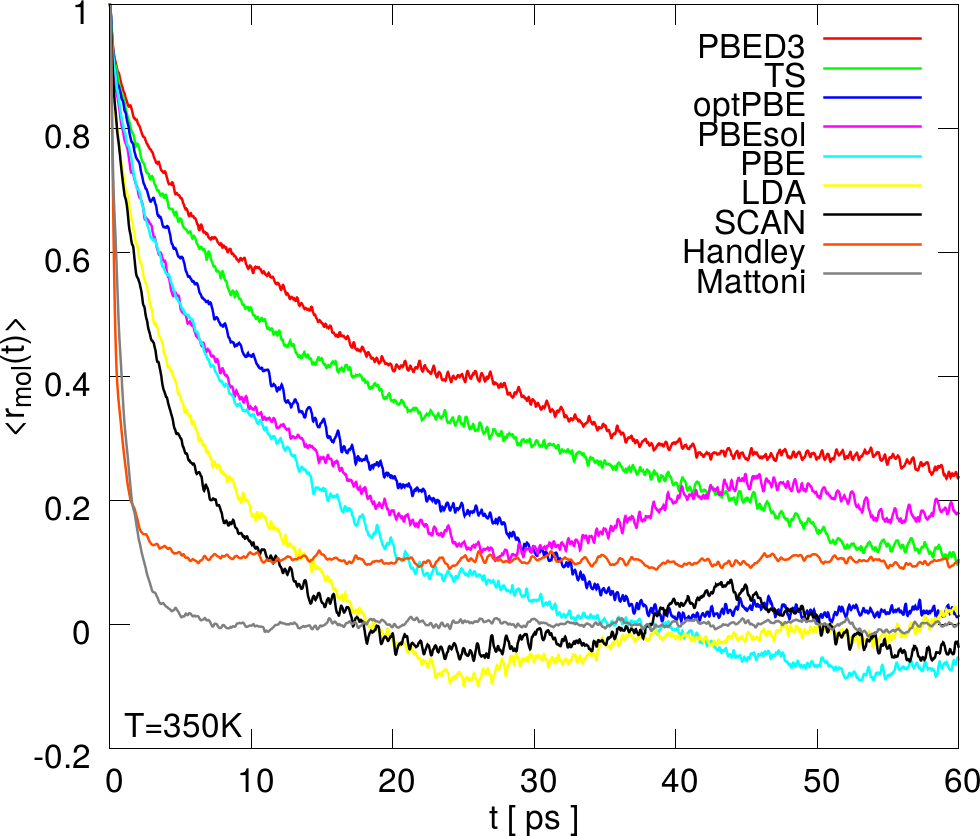}
 \caption{Autocorrelation functions $\left < r( t ) \right >$ of the MA molecules at $350$~K for 
different DFAs and force fields.}
 \label{self_corr_graphs}
\end{figure}


\begin{figure}[h!]
 \centering
 \includegraphics[width=0.45\textwidth]{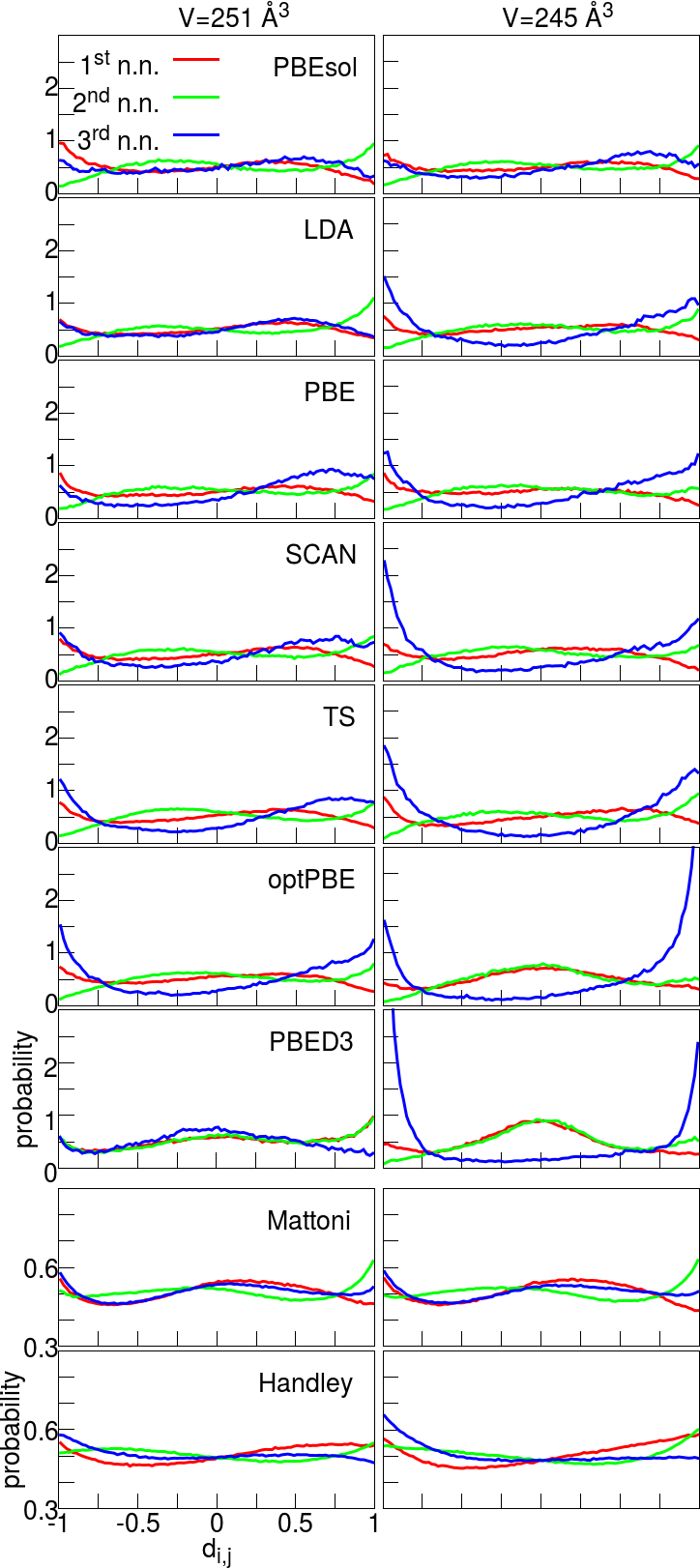}
 \caption{Probability distributions of relative molecular orientations ($d_{i,j}$) at 350~K. The left column corresponds to a unit cell volume of 251$~\text{\AA}^{3}$ and the right column to 245~$\text{\AA}^{3}$. Three different nearest neighbor shells are shown: 1$^{st}$ (red), 2$^{nd}$ (green) and 3$^{rd}$ (blue). Note the different scales for the probability axes for the force fields compared to the DFAs.}
 \label{relative_O}
\end{figure}

\subsection{Reorientation times of the MA molecules}
In Figure~\ref{self_corr_graphs}, the autocorrelation 
functions $\left < r( t ) \right >$ calculated from continues MD trajectories are shown. The reorientation times were determined by fitting $e^{-\nicefrac{t}{\tau{}}}$ to the autocorrelation functions as described in Section~\ref{sec:ana}. Table~\ref{timesbar} shows the reorientation times for the different functionals and temperatures. The stars denote that the MD simulation became trapped in a local minimum and therefore
the reorientation time can not be determined. For a free rotator with a fixed center of mass, the reorientation time $\tau$ scales with the temperature as $\nicefrac{1}{\sqrt{T}}$. As expected, the reorientation times decrease with increasing temperature, but do not scale as for the free rotator. Only for the Handley force field, this model results in a fit with a low rms error, consistent with the previously observed flat potential energy surface. Some irregularities are observed in the PBE and TS reorientation times. One possibility is that the randomized starting configurations influenced the reorientation time. This can occur even for reasonably long MD trajectories of a $2\times2\times2$ supercell, see for more details Ref.~\onlinecite{Lahnsteiner:prb16}. The system can become (temporarily) trapped in a phase where the 4-fold rotation about an axis perpendicular to the C-N bond is suppressed.

\subsection{Relative orientation of neighboring molecules}
In the cubic MAPbI$_3$ crystal, every MA molecule is surrounded by 6 nearest 
neighbors (n.n.) molecules, 12 second n.n. and 8 third n.n. Figure \ref{relative_O} shows the relative ordering of the molecules with respect to the neighboring molecules ($d_{i,j}$) in the different shells for the two different volumes. The distributions are calculated using Eq.~(\ref{rel_o}) at 350~K. A parallel/anti-parallel orientation of two molecules yields $d=1$/$-1$, respectively, and an orthogonal configuration yields $d=0$. In all graphs, the red line shows the ordering with respect to the first n.n., the green line to the second n.n. and the blue line to the third n.n. shell.
Extending the simulation time indicates that the results are well converged. Note that there is only one third n.n. for every MA molecule in the  $2\times2\times2$ super cell, therefore the distribution shows more noise.

In the calculations with the pseudo-cubic volume of 251~$\text{\AA}^{3}$, a trend in the relative ordering patterns is observed. This trend has the same order of the functionals as in the effective potential energy surfaces. Both PBEsol and LDA show an alternating pattern for the different n.n. shells. The first and third n.n. distributions are, within the statistical uncertainty, the same and the second is a mirror image of the first. The peaks in the distribution for the first and third n.n. correspond to an anti-aligned or an $60^{\rm o}$ configuration. The peaks for the second n.n. are at a $140^{\rm o}$ or an aligned
configuration. The distributions for PBE, SCAN and TS look similar, but the third nearest neighbor shows a small deviation from the more perfect alternating n.n. probability distribution. PBED3 and optPBE show a different relative ordering and do not show an alternating behavior at all. This is consistent with the potential energy surface which shows a different symmetry. The distributions of the different shells are almost the same and show, compared to the other functionals, a strong preference for orthogonal relative orientations. 

When the volume of the cell is reduced to 245~$\text{\AA}^{3}$ only small changes are observed in the distributions for the first and second n.n. shell. However, the ordering in third n.n. shell is strongly affected. In the most volume sensitive case, PBED3, the third n.n. order almost exclusively parallel and anti-parallel, while the first and second n.n. order orthogonal to the central MA molecule. Its potential energy surface shows that the eight molecules in the supercell prefer to occupy the six $\pm{}x$, $\pm{}y$, $\pm{}z$ axes orientations with equal probability. Since the barriers between these states are high at this volume the system is frustrated. The small volume reduction has almost no effect on the force fields. The distributions are more uniform compared to the DFAs, in agreement with the calculated "flat" potential energy surfaces.

\subsection{Iodine-hydrogen bond analysis}
\begin{figure}[t]
 \centering
 \includegraphics[width=1\linewidth]{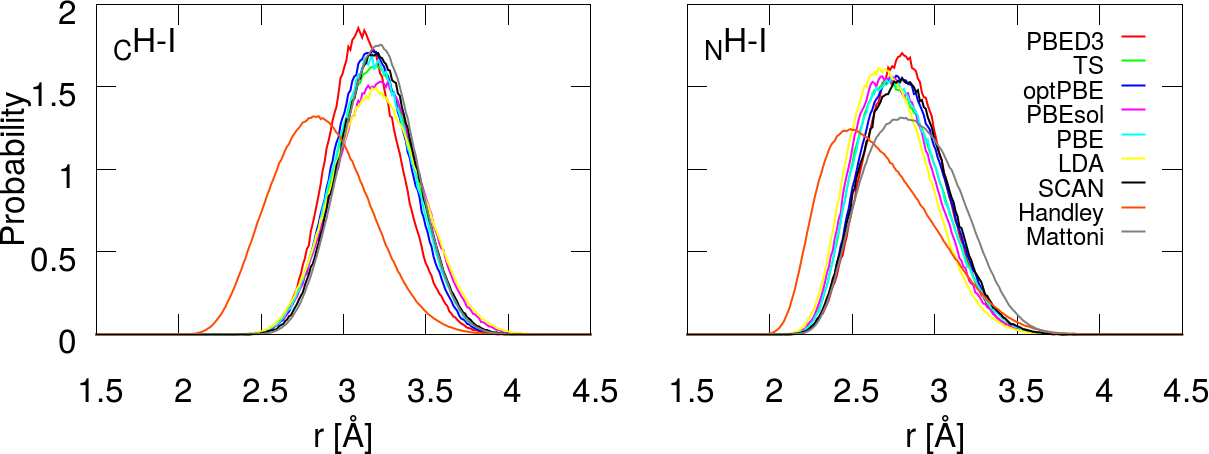}
 \caption{Probability distributions for the shortest \ce{_{C}H-I} (left) and 
\ce{_{N}H-I} (right) bonds at $350$~K for  various XC-functionals and force fields.}
 \label{HIbonds_fig}
\end{figure}

The lack of chemical bonding of the $\text{MA}^{+}$ molecule to the PbI framework makes that iodine-hydrogen bonds have an important influence on the molecular order\cite{Egger:jpcl14,Lee:sr16,Jingrui:prb16}. The hydrogen atoms of the \ce{CH_{3}-NH_{3}^{+}} molecules can bond weakly to iodine atoms. To analyze these bonds at finite temperature, the hydrogens are assigned to two groups throughout the MD trajectory. The first group contains the carbon-hydrogens (\ce{_{C}H}) and the second 
nitrogen-hydrogens (\ce{_{N}H}). Next the distance of each hydrogen is computed to the closest iodine atom. The three shortest \ce{_{N}H-I} and \ce{_{C}H-I} distances are then assigned to histograms shown in 
Fig.~\ref{HIbonds_fig}. An approximately Gaussian probability distribution for the iodine-hydrogen bonding lengths is found. The average value and the variance of those distributions are shown in Table~\ref{HI_bonds}. The difference of the average values between the different functionals is very small and there is no clear trend visible with respect to the functionals. This also holds for the variance. For all functionals the \ce{_{N}H-I} bonding length is increasing with temperature whereas the \ce{_{C}H-I} is decreasing with temperature. This indicates that the molecules connect with the nitrogen side to the iodine atoms via hydrogen bonding. At higher temperatures the hydrogen bonding weakens and the molecules will move more to the center of the PbI cages. The \ce{_{N}H} atoms remain closer to the iodines in agreement with the higher electronegativity of the nitrogen atom. The force fields are able to reproduce the temperature trend suggested by the ab initio simulations.  However, the average values of the Handley force fields are significantly shorter and the Mattoni force field shows too long \ce{_{N}H-I} bonds.

\begin{table*}
 \caption{\ce{_{N}H-I} and \ce{_{C}H-I} hydrogen bond lengths and their variances for 
  various functionals and temperatures. Units for bond lengths is \AA \ and
  the variance ($\sigma^2$) in units of $\text{\AA}^{2}$.}
  \footnotesize
  \begin{ruledtabular}
  \begin{tabular}{|l|c|c|c|c|c|c|c|c|c|c|c|c|c|c|c|c|c|c|c|c|c|}
  & T    &   \multicolumn{2}{c|}{PBEsol}  & \multicolumn{2}{c|}{LDA} &  \multicolumn{2}{c|}{PBE} &  \multicolumn{2}{c|}{SCAN} &  \multicolumn{2}{c|}{TS} &  \multicolumn{2}{c|}{optPBE} &  \multicolumn{2}{c|}{PBED3} &\multicolumn{2}{c|}{Mattoni}& \multicolumn{2}{c|}{Handley} &exp.\cite{Franz:crat2016}&exp.\cite{Weller:cc105}  \\

  &    &  mean &$\sigma^2$&  mean &$\sigma^2$& mean &$\sigma^2$& mean&  $\sigma^2$&  mean &$\sigma^2$&   mean &$\sigma^2$ &mean &$\sigma^2$ & mean & $\sigma^2$ & mean & $\sigma^2$ & mean& mean\\
  \hline
 \ce{_{N}H-I}& 250~K &  2.75 &  0.05 & 2.72  &  0.05 & 2.75 &  0.05 & 2.79  &  0.05 & 2.79  &  0.05 &  2.78  &  0.05 & 2.81  &  0.04   & 2.83 & 0.07 & 2.63 & 0.09    & &    \\
             & 300~K &  2.76 &  0.05 & 2.73  &  0.05 & 2.77 &  0.06 & 2.81  &  0.06 & 2.81  &  0.06 &  2.81  &  0.05 & 2.81  &  0.05                & 2.85 & 0.07 & 2.65 & 0.09    & &    \\
             & 350~K &  2.77 &  0.06 & 2.74  &  0.06 & 2.79 &  0.06 & 2.82  &  0.06 & 2.82  &  0.05 &  2.81  &  0.06 & 2.82  &  0.05                & 2.87 & 0.07 & 2.68 & 0.10    & 2.84 & 3.12   \\
             & 400~K &  2.79 &  0.07 & 2.75  &  0.06 & 2.80 &  0.06 & 2.83  &  0.06 & 2.82  &  0.06 &  2.83  &  0.06 & 2.82  &  0.06                & 2.89 & 0.08 & 2.69 & 0.10    & &\\
\hline
 \ce{_{C}H-I}& 250~K &  3.24  & 0.05 &  3.24 &  0.06 &  3.21 &  0.05&  3.20  &  0.05 &3.23  &  0.04&  3.19  &  0.04 & 3.15&  0.04 & 3.23 & 0.04 & 2.87 & 0.07   & &    \\
             & 300~K &  3.22  & 0.06 &  3.22 &  0.06 &  3.19 &  0.05&  3.20  &  0.05 &3.22  &  0.05&  3.18  &  0.05 & 3.14&  0.04             & 3.22 & 0.05 & 2.87 & 0.08   & &    \\
             & 350~K &  3.21  & 0.06 &  3.21 &  0.07 &  3.18 &  0.05&  3.19  &  0.05 &3.20  &  0.05&  3.17  &  0.05 & 3.13&  0.05             & 3.21 & 0.05 & 2.86 & 0.08   & 3.09 & -  \\
             & 400~K &  3.19  & 0.07 &  3.19 &  0.07 &  3.18 &  0.06&  3.19  &  0.06 &3.19  &  0.06&  3.16  &  0.05 & 3.13&  0.05              & 3.20 & 0.05 & 2.85 & 0.09   & &

 \end{tabular}
 \end{ruledtabular}
 \label{HI_bonds}
\end{table*}


\section{Discussion}
\label{sec:dis}

In this section the structural parameters calculated with the different computational methods are compared to available experimental data. We study which calculated structural properties can act as effective selection criteria to determine the most accurate computational method(s) for MAPbI$_3$.

\subsection{Structure: PbI$_3$ framework and MA orientations}

From many experiments, it is known that the perovskite has
cubic symmetry above $\sim330$~K and that the framework (at least on time-average)
should be aligned with the principal axes\cite{Stoumpos:ic13,Baikie:jmca:13,Weller:cc105,Whitfield:sr16,Franz:crat2016}. This means
the orientational distributions of the \ce{I-I} vectors in the framework should
resemble symmetrically smeared spots, where the radius of smearing indicates the stiffness of the framework. The 
tetragonal phase is stable from $\sim160$~K to $\sim330$~K and
the PbI octahedra show clockwise-counterclockwise rotations resembling a PbI zigzag pattern in the $xy$-plane\cite{Whitfield:sr16,Weller:cc105,Franz:crat2016}. In the probability 
distributions for the framework shown in this work, this agrees
with spots left and right of the $y$-direction. Below $\sim160$~K the perovskite possesses the orthorhombic structure which shows a 3-dimensional zigzag pattern in and out of the $xy$-plane \cite{Whitfield:sr16,Weller:cc105}. In the framework probability distributions this would result in
4 spots distributed around $y$ direction. These three typical structural characteristics are shown in Fig.~\ref{frame_polar_sketch}(b). 

In all calculations the volume of the supercell has been fixed at the experimental volume at 400~K. When comparing this description to the obtained I-I distribution functions of Fig.~\ref{frame_polars_T}, it is clear that at $350$~K LDA, PBE, PBEsol, SCAN, TS, optPBE and the Mattoni force field describe the perovskite framework in agreement with experiment. However, LDA, PBEsol, TS and the Mattoni force field do not show a sign for a transition to a different phase when the temperature is decreased. The optPBE functional does show a sign, however a pattern resembling the orthorhombic phase starts building already at $250$~K. The remaining functionals, SCAN and PBE, are able to describe the cubic phase but also show the zigzag pattern of the PbI octahedra in the tetragonal phase when the temperature is lowered. According to experiment the tetragonal phase should already be observed in the simulations at $300$~K. However, since the lattice constants are kept fixed during all our simulations, the onset of the cubic to tetragonal phase transition is likely to be shifted in temperature.

\begin{figure}[!b]
 \centering
 \includegraphics[width=0.4\linewidth,center]{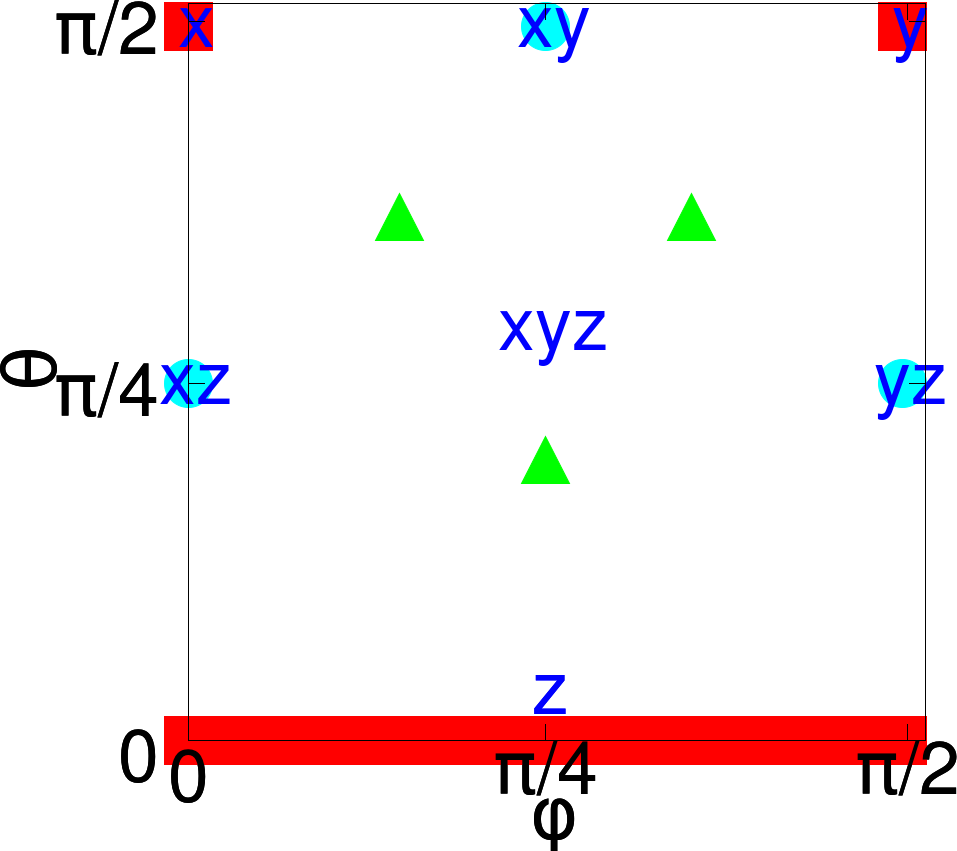}
 \caption{Experimental orientations found by Whitfield et al.~\cite{Whitfield:sr16} and Weller et al.~\cite{Weller:cc105} (red squares), Franz et al.~\cite{Franz:crat2016} 
(green triangles) and Kawamura et al.~\cite{Kawamura:jpsj01}(turquoise circles). The labels indicate the high symmetry directions in the reduced polar distribution.}
\label{exp_orient}
\end{figure}

The experimentally observed $\text{MA}^{+}$ orientations in the high temperature ($\sim350$~K) 
pseudo-cubic phase from three different sources are illustrated in the polar plot of Figure~\ref{exp_orient}. In the work of Franz~\textit{et al.} the molecules are shown to have a preference for a tilted of room diagonal orientation. Non of the here shown methods show a preference for this orientation. At the same temperature, Whitfield~\textit{et al.}~\cite{Whitfield:sr16} and Weller~\textit{et 
al.}~\cite{Weller:cc105} observed that the molecules preferentially align along the Cartesian 
axes. For the simulations done at 251~\AA$^{3}$ only PBE-D3, optPBE and the Mattoni force field
are in agreement with this experimental observation. By decreasing the unit cell volume to
245~\AA$^{3}$ also SCAN and TS are preferentially aligning the molecules along the principal axes. And Kawamura et al.~\cite{Kawamura:jpsj01} observed molecules orienting
along the face diagonals for the cubic phase. This is in agreement with the results obtained by the non van der Waals DFAs for the simulations in the cubic volume.

The large spread in the experimental data suggest that a common experimental consensus is 
still missing.

\subsection{Energy and time related to MA reorientation}
According to the here used definition of the maximal corrugation energy ($\Delta{}E$) it is an upper bound to the rotational barrier for the molecules. The barrier is a statistically averaged rotation of the molecule over all pathways on the energy surface, connecting the initial with the final state, that are accessible at the considered temperature. However, this is not the only complicating factor, when comparing rotational barriers measured in experiment (see Table~\ref{T1:exp_data}) with the barrier heights calculated in this work (see Table~\ref{timesbar}). The 
spread in the values among the different experiments is much larger than the 
spread for the different DFAs. This is not only due to the 
different experimental and sample conditions, but also due to the different 
underlying models used to extract a barrier from the experimentally recorded 
signals. All of the computational methods show corrugation energies (see Table~\ref{T1:exp_data}) of roughly the same magnitude. Only PBE-D3 functional and the Handley force field can be considered to be outliers, because the corrugation energies are too large or too low, respectively. A very similar and closely related problem appears 
when comparing the reorientation times ($\tau$) of the molecules. In the experimental data for 
the tetragonal structure at 300~K, a spread in the data from $0.5$~ps to 
$14$~ps is reported. The spread in the reorientation times obtained from the MD calculations at 
$350$~K is of the same order. By comparing the MD data 
with the experimental data at $350$~K, we see that the PBED3 functional (19~ps) differs the most from the experimental value of $2.7$~ps. By comparison, SCAN fits best to the 
experimental reorientation time. 

The experimentally measured hydrogen-iodine bond lengths (\ce{_{N}H-I}, \ce{_{C}H-I}) are included in Table~\ref{HI_bonds}. Comparing with the experimental work of Franz~\textit{et al.}\cite{Franz:crat2016} at 350~K, all density functionals and the Mattoni force field yield shorter \ce{_{N}H-I} and larger \ce{_{C}H-I} bond lengths. Weller~\textit{et al.}~\cite{Weller:cc105} reports a change in the \ce{_{N}H-I} bond lengths depending on the temperature: 2.61-2.81 (100~K), 3.15-3.18 (180~K) and 3.12-3.52 (352~K). The spread between the experimental data at 350~K is too large to make a best choice between the methods. The Handley force field does not give a correct description of the hydrogen bonding in the perovskite. According to the 0~K DFT structure optimization work of Li~\textit{et al.} hydrogen bonding is well described by PBE in hybrid perovskites\cite{Jingrui:prb16}. All methods show approximately the same increase in \ce{_{N}H-I} as function of temperature as PBE, except for PBED3. We speculate that the thermal motion of the PbI framework and the MA molecule softens the hydrogen bonding and this contributes to an effective potential surface which allows reorientation of the MA molecules even at room temperature.


\section{Summary and Conclusions}
\label{sec:conc}
A comparison of \textit{ab intio} and classical molecular dynamics methods applied to the hybrid perovskite MAPbI$_3$ was presented. Based on one consistent structural analysis, various density functional approximations (DFAs) and force fields were compared to experimental data. Currently, the spread in the experimental data is too large to uniquely determine the most accurate computational approach for this highly dynamic ionic crystal. However, promising candidates and outliers have been identified. 

The PbI octahedra create a cavity for the molecules to move in. By calculating the distribution function of the PbI octahedra orientations at temperatures between 250~K and 400~K, we have shown that the PBE and SCAN functionals show changes in the structure that can well be related to the experimentally observed
phase transition from the cubic to the tetragonal structure. For other DFAs, the evidence is less conclusive. PBEsol and LDA show no clear cut structural change upon cooling, whereas for the optPBE and PBE-D3 (and to some extent TS) the order increases, but the observed ordering rather points to an orthorhombic like instead of tetragonal  order. As eluded to in the main text, there are caveats to this statement. Since tetragonally distorted unit cells have not been considered, and since the simulation times are still short---
albeit quite long by usual first principles standards --- we are left with somewhat unambiguous results. In future, one would certainly like to reinspect the present results with larger unit cells, even longer simulation times, and variable cell shape simulations. Whether this will be feasible  with first principle techniques in the near future remains, however, somewhat questionable.

Concerning the orientation of the MA molecules we have made the following observations.
The effective potential for the molecular orientation (obtained by Boltzman inversion) 
can be accurately modeled by a linear combination of only two cubical harmonics. In the cubic phase at 350~K, all here considered DFAs and force fields show an effective potential energy surface and reorientation time for the MA molecules consistent with rotational freedom. 
The maximal corrugation energy lies between 30-50~meV, and the reorientation time lies between 4-14~ps. The van der Waals corrected PBE-D3 functional is the only exception yielding 76~meV and 19~ps. This seems fairly large and not quite consistent with most experimental findings. 
All DFAs agree that the molecules avoid a C-N orientation along the room diagonal. Whether the face diagonal or the principle axes are preferred, however, depends on the functional. In the high temperature cubic phase, the simple local and semi-local functionals prefer orientations along the face diagonals, whereas optPBE and PBE-D3 orient the molecules preferably along the principal axes. 

A greatly troublesome outcome of our work is that the volume of the simulation cell can quite drastically change the results. In the present work, we have used two volumes for our simulations, the experimental volume of the high temperature cubic phase and the optimal volume of the tetragonal phase as predicted by the SCAN functional\cite{Bokdam:prl17}; this corresponds to a 2~\% volume change. For the SCAN and optPBE functional, this 2~\% volume change changes the preferred C-N orientation of the molecules from along the principle axes to along the face diagonal. These observed changes upon a small volume decrease clearly imply that significant attention needs to be paid to the considered volumes. It also means that even in the experiment the molecules might change their preferred orientation between the lower volume tetragonal phase and the larger volume cubic phase. That such a small change of the volume can have such a drastic effect on the order of the molecules is 
unheard of, and requires a careful re-evaluation of simulations performed in the past.  

Returning to the molecular order, we have also investigated the relative order of neighboring molecules. In particular at higher temperature (350 K), the molecular orientations are fairly random with a modest long range order. At the lower volume, the PBE-D3 and optPBE functionals lead to a more pronounced long range order than functionals without vdW corrections. This is in line with what one would expect and correlates with the slower reorientation time of the molecules for vdW functionals.

Finally, let us briefly comment on the available classical force fields. Neither the Mattoni~\textit{et al.} nor the Handley \textit{et al.} force fields yield results that are close to any of the considered DFA functionals. Both yield a very rigid PbI framework. Additionally, the  Mattoni force field shows (in the here chosen setup of fixed lattice vectors) no sign of a phase transition; the PbI bonds are always aligned to the principle axis. The Handley force field does yield a transition, most likely to the orthorhombic phase around 350~K. For the orientation of the MA molecule, the Handley force field shows no preferred order and resultantly a free rotator behavior with far too short reorientation times. The Mattoni force field performs better, but the reorientation speed is still too fast compared to the DFAs. Clearly the present force fields for MAPbI$_3$ are not satisfactory, and improvements are called for. 

Taking all things together, our conclusion is that the PBE and SCAN functionals are showing a very reasonable performance in line with the experimental data. However, we have given PBE a somewhat unfair ``bonus'' here:
the simulations were done at the experimental volume, although PBE predicts far too large volumes. 
Given the large volume sensitivity discussed above, we can not predict how PBE would perform had we adopted
the theoretically optimized volume. This leaves SCAN as the best choice for MD simulations, and
we recall that the SCAN functional predicts a tetragonal instability at temperatures close to
the experimentally observed cubic to tetragonal transition temperature\cite{Bokdam:prl17}.
A final word of caution needs to be made concerning the PBE-D3 functional. As clearly shown in Fig.~\ref{e_surf_orig_350K_fit}, the molecules are very immobile even at 350~K in the cubic structure. Also there is hardly any structural change for the PbI framework between 400~K and 250~K. We feel that
this is not in agreement with the experimental data, but due to the large spread in the outcomes of the different experimental works, it is difficult to make a final judgment.

\acknowledgements{We would like to thank David Dubbeldam for valuable input for the classical simulations with RASPA. Funding by the Austrian Science Fund (FWF): P 30316-N27 is gratefully acknowledged. Computations were performed on the Vienna Scientific Cluster VSC3. }

%


\section{Appendix}

\begin{table*}
 \caption{Pearson-correlation coefficients of the raw MD potential energy surface data and fitted model 
function at a pseudo-cubic volume of 251~$\text{\AA}^{3}$.}
 \label{corr_coeffs1}
\begin{ruledtabular}
 \begin{tabular}{c c c c c c c c c c }
    T &  PBEsol &  LDA &  PBE &  SCAN &  TS &  optPBE &  PBED3 &  Mattoni &  Handley \\ 
 250~K &  0.91 &  0.93 &  0.91 &  0.92 &  0.93 &  0.97 &  0.98 &  0.93 &  0.74 \\ 
 300~K &  0.92 &  0.96 &  0.92 &  0.91 &  0.96 &  0.98 &  0.97 &  0.94 &  0.84 \\ 
 350~K &  0.88 &  0.92 &  0.87 &  0.87 &  0.97 &  0.97 &  0.98 &  0.96 &  0.80 \\ 
 400~K &  0.86 &  0.94 &  0.89 &  0.85 &  0.96 &  0.96 &  0.98 &  0.97 &  0.90 \\ 
 \end{tabular}
 \end{ruledtabular}
\end{table*}

\begin{table*}
 \caption{Pearson-correlation coefficients of the raw MD potential energy surface data and fitted model 
function at a pseudo-cubic volume of 245~$\text{\AA}^{3}$.}
 \label{corr_coeffs2}
\begin{ruledtabular}
 \begin{tabular}{c c c c c c c c c c }
   T         &   PBEsol      &    LDA    &  PBE    &  SCAN   &   TS   &  optPBE  &  PBED3     &  Mattoni  &  Handley  \\ 
  350~K     &   0.85    &    0.87   &   0.95  &  0.94   &   0.96  &   0.98   &    0.97    &  0.96   &   0.87  \\ 
   
 \end{tabular}
 \end{ruledtabular}
\end{table*}
\begin{table*}
 \caption{Fitted parameters $C_1$, $C_2$ and $C_{3}$ (in meV) describing the 
potential energy surface for the different DFAs and force fields at a pseudo-cubic volume of 251~$\text{\AA}^{3}$. }
 \begin{ruledtabular}
 \begin{tabular}{c c c c c c c c c c c c c c c c }
   T   & \multicolumn{3}{c}{PBEsol} & \multicolumn{3}{c}{LDA} & \multicolumn{3}{c}{PBE} & \multicolumn{3}{c}{SCAN} & \multicolumn{3}{c}{} \\ 
                                                                    
  250~K &       -68.8  &  1520.6  &  17.0  &  -84.1  &  1494.5  &  20.9  &  -54.7  &  1620.7  &  13.5  &  -44.2   &  1404.9  &  10.9  \\ 
  300~K &       -72.9  &  1523.7  &  18.1  &  -85.2  &  1493.1  &  21.1  &  -71.9  &  1616.3  &  17.8  &  -45.7   &  1219.8  &  18.0  \\ 
  350~K &       -72.7  &  1336.2  &  18.0  &  -63.8  &  1320.7  &  15.8  &  -70.4  &  1435.7  &  17.43  &   -38.2   &  1035.4  &  9.4  \\ 
  400~K &       -71.2  &  1307.8  &  17.7   &  -72.7  &  1323.4  &  18.0  &  -59.4  &  1343.7  &  14.7  &  -44.7   &  1024.7  &  11.1  \\ 
   & & & & & & & & & & & & & & & \\
       & \multicolumn{3}{c}{TS} & \multicolumn{3}{c}{optPBE} & \multicolumn{3}{c}{PBED3} & \multicolumn{3}{c}{Mattoni} & \multicolumn{3}{c}{Handley} \\
  
  250~K &  -48.6   &  1765.3  &  12.0  &  78.6  &  1108.5  &  -0.1  &  224.1  &  93.4  &  -0.2  &   68.9  &  417.9  &  -0.1  &  -74.7  &  369.1  &  18.6 \\ 
 
  300~K &  -43.5   &  1708.1  &   10.7  &  79.8   &   960.2  &  -0.1  &  192.2  &  282.8  &  -0.2  &  77.1  &  211.0  &  -0.1  &  -65.3  &  235.1  &  16.3 \\  
  350~K &  -40.0   &  1452.1  &  9.8  &  57.5   &  876.0   &  -0.1  &  222.4  &  44.0   &  -0.2  &   92.7  &  76.3  &  -0.1   &  -49.8  &  147.2  &  12.5 \\ 
  400~K &  -46.8   &  1380.2  &  11.5  &  51.2   &  735.2   &  -0.1  &  196.7  &  268.3  &  -0.2  &  102.2  &  4.5    &  -0.1  &  -50.9  &  84.1  &  13.9 \\  
       & & & & & & & & & & & & & & & \\                                                            
      & $C_1$  &$C_2$  &$C_3$ &$C_1$  &$C_2$&$C_3$ &$C_1$ &$C_2$ &$C_3$ &$C_1$&$C_2$& $C_3$ &$C_1$&$C_2$& $C_3$\\  
 \end{tabular}
 \end{ruledtabular}
 \label{Tfit1}
\end{table*}

\begin{table*}
 \caption{Fitted parameters $C_1$, $C_2$ and $C_{3}$ (in meV) describing the 
potential energy surface for the different DFAs and force fields at a pseudo-cubic volume of 245~$\text{\AA}^{3}$. }
 \begin{ruledtabular}
 \begin{tabular}{c c c c c c c c c c c c c c c c}
   T   & \multicolumn{3}{c}{PBEsol} & \multicolumn{3}{c}{LDA} & \multicolumn{3}{c}{PBE} & \multicolumn{3}{c}{SCAN} & \multicolumn{3}{c}{} \\           
  350~K &   -60.6  &  1154.5  &  15.0  &  14.0  &  1091.3  &  0.0  &  29.7  &  1123.4  &  0.0  &  100.0   &  629.6  &  -0.1 & & \\
     & & & & & & & & & & & & & & & \\                                                            
  & \multicolumn{3}{c}{TS} & \multicolumn{3}{c}{optPBE} & \multicolumn{3}{c}{PBED3} & \multicolumn{3}{c}{Mattoni} & \multicolumn{3}{c}{Handley} \\
  350~K &  91.5   &   1047.6  &  -0.1  &   230.3   &  126.8   &  -0.2  &  345.6  &  -257.0  &  -0.3  &  75.6   &  231.8   &  -0.1  &  -51.8  &  75.0  & 14.5 \\
       & & & & & & & & & & & & & & & \\                                                            
      & $C_1$  &$C_2$  &$C_3$ &$C_1$  &$C_2$&$C_3$ &$C_1$ &$C_2$ &$C_3$ &$C_1$&$C_2$& $C_3$ &$C_1$&$C_2$& $C_3$\\ 
 \end{tabular}
 \end{ruledtabular}
 \label{Tfit2}
\end{table*}

\end{document}